\documentclass[12pt]{article}
\usepackage{amsmath,amssymb,times,graphicx}
\begin{document}
\pagestyle{plain}
\hsize = 6. in 				
\vsize = 8.5 in		
\hoffset = -0.3 in
\voffset = -0.5 in
\baselineskip = 0.26 in	

\def\vF{{\bf F}}
\def\vJ{{\bf J}}
\def\vX{{\bf X}}
\def\tf{{\widetilde{f}}}
\def\vv{{\bf{v}}}
\def\tv{{\widetilde{v}}}
\def\tu{{\widetilde{u}}}
\def\tp{{\widetilde{\rho}}}
\def\tR{{\widetilde{R}}}
\def\vx{\mbox{\boldmath$x$}}

\title{A Decomposition of Irreversible
Diffusion Processes Without Detailed Balance}

\author{Hong Qian\\[10pt]
Department of Applied Mathematics\\
University of Washington, Seattle\\
WA 98195-2420, U.S.A.}

\maketitle

\begin{abstract}
As a generalization of deterministic, nonlinear conservative
dynamical systems, a notion of {\em canonical conservative dynamics}
with respect to a positive, differentiable stationary 
density $\rho(x)$ is introduced:
$\dot{x}=j(x)$ in which $\nabla\cdot\big(\rho(x)j(x)\big)=0$.
Such systems have a conserved ``generalized free energy
function'' $F[u]$ $=$ $\int u(x,t)\ln\big(u(x,t)/\rho(x)\big)dx$
in phase space with a density flow $u(x,t)$ satisfying
$\partial u_t =-\nabla\cdot(ju)$.  Any general stochastic
diffusion process without detailed balance, in terms of its
Fokker-Planck equation, can be decomposed into a 
reversible diffusion process with detailed balance and 
a canonical conservative dynamics.  This decomposition can be 
rigorously established in a function space with inner 
product defined as 
$\langle\phi,\psi\rangle=\int\rho^{-1}(x)\phi(x)\psi(x)dx$. 
Furthermore, a law for balancing $F[u]$ can be 
obtained: The non-positive 
$dF[u(x,t)]/dt$ $=$ $E_{in}(t)-e_p(t)$ 
where the ``source'' $E_{in}(t)\ge 0$ and the
``sink'' $e_p(t)\ge 0$ are known as 
house-keeping heat and entropy production, respectively.  
A reversible diffusion has $E_{in}(t)=0$.
For a linear (Ornstein-Uhlenbeck) diffusion process, our 
decomposition is equivalent to the previous approaches 
developed by R. Graham and P. Ao, as well as the theory 
of large deviations.  In terms of  two different formulations
of time reversal for a same stochastic process,
the meanings of {\em dissipative} and {\em conservative}
stationary dynamics are discussed.
\end{abstract}

\section{Introduction}

	With the recent development of stochastic thermodynamics in terms 
of mesoscopic entropy production \cite{esposito,seifert,crooks99,
qqg91}, free energy dissipation \cite{santillan-qian,ge_qian_10,3faces}, 
work equalities and fluctuation theorems \cite{jarzyn,sasa01,
lebowitz-sphon,kurchan}, and the mathematical theory of nonequilibrium 
steady state \cite{jqq,zqq,gqq,qq85}, there is a revitalized 
interest in nonlinear stochastic dynamics \cite{qian_qb,qian_nonl}, 
particularly those without detailed balance \cite{ge_qian_chaos,ao_ctp}. 

	Nonlinear stochastic dynamics without detailed balance can
be mathematically represented by irreversible Markov processes.   
For a general discrete state Markov process, either with 
discrete or continuous time parameter, a decomposition theorem is
known \cite{qq82,jqq,qqg91}.  In a nutshell, the transition
probability matrix of a Markov process, with respect to its
stationary distribution, can be decomposed in terms of 
a symmetric and an anti-symmetric parts.  The latter part
can be further decomposed into many pure rotations among
the discrete states.  The notion of cycle kinetics arises in
this analysis \cite{kalpazidou,hill,schnakenberg}. 

	For continuous diffusion processes on $\mathbb{R}^n$ 
without detailed balance, such a decomposition has not 
been fully established, even though computations have revealed 
both stationary density and rotational flux as key determinants 
of a stationary process \cite{wang_pnas_08,feng_jcp_11}. 
A sophisticated analysis on a compact differentiable 
manifold with genus also exists \cite{jqq,qwcmp}.  
In the physics literature, R. Graham and coworkers have proposed
and studied extensively a decomposition of non-gradient 
vector field in terms of Fokker-Planck equations, via 
WKB method and a Hamilton-Jacobi equation \cite{graham_71,graham_73,graham_89}. But 
the program was ultimately abandoned due 
to technical difficulties \cite{graham_tel}.  In
recent years, P. Ao and coworkers have again proposed a 
related decomposition from a rather different starting
point \cite{kat,ao_jpa_04,ao_jpa_06,aoping_qian,ao_ctp}.  However, 
the feasibility of this new approach has been 
rigorously demonstrated only for a linear system \cite{kat} 
which is nearly equivalent, apart from the 
normalizability of the invariant density, to analyzing
irreversible Ornstein-Uhlenbeck processes \cite{qian_prsa}.  Still, their emphasis on 
stable as well as unstable fixed points had
suggested the possible applicability to nonlinear systems. A full analysis beyond heuristic for general nonlinear diffusion processes without detailed balance still is not available.  See \cite{ao_jpa_06,shi,qian_2nd} for more, and recent discussions.

	The centrepiece of the Fokker-Planck equation of a 
diffusion process is a linear, partial differential 
operator $\mathcal{L}$ \cite{qqt_jsp,jqq}.  Assuming the
existence of a unique stationary density $\rho(x)>0$, $x\in\mathbb{R}^n$ and $\mathcal{L}(\rho)=0$, an inner 
product in a function space 
\begin{equation}
     \langle\phi,\psi\rangle =  \int_{\mathbb{R}^n} 
               \phi(x)\psi(x)\rho^{-1}(x) dx
\label{innerprod}
\end{equation}
can be introduced \cite{qqt_jsp,jqq,ge_qian_chaos}.  For 
diffusion process with detailed
balance, this functional analysis approach is reduced to the
Sturm-Liouville problem.  In the present work, we follow this
approach and introduce an operator decomposition in terms
of a symmetric $\mathcal{L}_s$ and an anti-symmetric
$\mathcal{L}_a$: $\mathcal{L}=\mathcal{L}_s+\mathcal{L}_a$. 
The partial differential equation (PDE) 
$\big(\partial/\partial t
-\mathcal{L}_s\big)(u)=0$ is a parabolic PDE with a self-adjoint 
elliptic operator $\mathcal{L}_s$, representing a reversible diffusion
process and a gradient-like system, as expected for a symmetric
operator.  In sharp contrast, the 
PDE $\big(\partial/\partial t -\mathcal{L}_a\big)(u)=0$
turns out to be a first-order hyperbolic PDE in which the 
$\mathcal{L}_a$ is only a first-order differential operator with 
respect to $x\in\mathbb{R}^n$. 

We show that the anti-symmetric system corresponds 
to a generalization of the conservative nonlinear dynamics one 
usually studies \cite{perko}, where the $\rho$ is 
uniform.  One important example of this class of 
dynamics is Hamiltonian systems. 
We shall call the usual conservative dynamics
$\dot{x}=g(x)$ with $\nabla\cdot g(x)=0$ 
{\em microcanonical-like}.  The terminology is 
borrowed from statistical physics where microcanonical
system is a Hamiltonian dynamics in phase space
with Liouville's equation \cite{khinchin}.
Then the dynamical system
defined by $\mathcal{L}_a$ with $\dot{x}=j(x)$ in which 
$\nabla\cdot\big(\rho(x)j(x)\big)=0$ could be called a
{\em canonical-like}.  Indeed, for the density function
$u(x,t)$ in the phase space, a microcanonical-like dynamics has 
\begin{subequations}
\begin{equation}
                  \frac{d}{dt}\left(-\int_{\mathbb{R}^n}
                   u(x,t)\ln u(x,t)\ dx \right) = 0,
\label{nond_1}
\end{equation}
a result known to Boltzmann \cite{dorfman}.  Similarly in a 
canonical-like dynamics we have (see below)
\begin{equation}
                  \frac{d}{dt}\left[\int_{\mathbb{R}^n}
            u(x,t)\ln\left(\frac{u(x,t)}{\rho(x)}\right) 
                 dx \right] = 0.
\label{nond_2}
\end{equation}
\end{subequations}
In the present work, Eqs. (\ref{nond_1}) and (\ref{nond_2}) are 
used as the definitions for microcanonical and canonical conservative 
dynamics, respectively.

\section{Fokker-Planck differential operator decomposition}

A general Fokker-Planck equation is characterized by a symmetric,
non-singular $n\times n$ diffusion matrix $A(x)$ and a vector field, 
$b(x)$, the drift.  In the present work, we shall denote the
linear differential operator in a function space
\begin{eqnarray}
          \mathcal{L}(\phi) &=&   
                 \nabla\left(A(x)\nabla\phi(x)\right)
                   -\nabla\left(b(x)\phi(x)\right),
\\
          \mathcal{L^*}(\phi) &=& \nabla\left(A(x)\nabla\phi(x)\right)
               +b(x)\nabla\phi(x),
\end{eqnarray}
and adopt the inner product given in Eq. (\ref{innerprod}),
in which $\rho(x)$ is the normalized stationary solution to 
the PDE $\mathcal{L}(\rho)=0$. We assume the $\rho(x)$ is unique, 
positive and differentiable \cite{qqt_jsp,jqq}.  Then we 
have
\begin{eqnarray*}
      \langle\psi,\mathcal{L}(\phi)\rangle 
        &=& \int_{\mathbb{R}^n} \rho^{-1}\psi(x)\ dx\ 
             \Big\{ \nabla\left(A(x)\nabla\phi(x)\right)
                   -\nabla\left(b(x)\phi(x)\right)\Big\} 
\\[8pt]
        &=& \int_{\mathbb{R}^n} 
             \Big\{\nabla\left[A(x)\nabla\left(\rho^{-1}
            \psi(x)\right)\right]+b(x)\nabla\left(\rho^{-1}
            \psi(x)\right) \Big\} \phi(x)dx
\\[8pt]
      &=& \langle\rho\mathcal{L}^*\left(\rho^{-1}\psi\right),
                   \phi\rangle.
\end{eqnarray*}

We now introduce symmetric and anti-symmetric operators
\begin{eqnarray}
        \mathcal{L}_s(\phi) &=&\frac{1}{2}\left\{
             \mathcal{L}(\phi) + \rho\mathcal{L^*}
            \left(\rho^{-1}\phi\right)\right\},
\\
         \mathcal{L}_a(\phi) &=& \frac{1}{2}\left\{
             \mathcal{L}(\phi) - \rho\mathcal{L^*}\left(\rho^{-1}
            \phi\right)\right\}.
\end{eqnarray}
Then it is easy to verify:
\begin{equation}
   \langle\psi,\mathcal{L}_s(\phi)\rangle
        =\frac{1}{2}\Big(\langle \psi,\mathcal{L}(\phi)\rangle 
         + \langle\mathcal{L}(\psi),\phi\rangle\Big)
        = \langle\mathcal{L}_s(\psi),\phi\rangle,
\end{equation}
\begin{equation}
   \langle\psi,\mathcal{L}_a(\phi)\rangle
        = -\langle\mathcal{L}_a(\psi),\phi\rangle,
\end{equation}
and $\mathcal{L}$ has a decomposition
$\left(\mathcal{L}_s+\mathcal{L}_a\right)$.

	In physics literature, one usually prefers to write 
Fokker-Planck equation.  However, in mathematical literature,
a diffusion process is represented by its infinitesimal generator
$\widetilde{\mathcal{L}}=\nabla A(x)\nabla + b(x)\nabla$.
Then one defines an inner product with respect to the 
invariant measure $\rho(x)$ as $\langle u,v\rangle^{\rho}$ $=$
$\int_{\mathbb{R}^n} u(x)v(x)\rho(x)dx$.  With respect to
$\rho(x)$, the $\widetilde{\mathcal{L}}$ $=$
$\rho^{-1}(x)\nabla\big( A(x)\rho(x)\nabla\big)$ $+j(x)\nabla$
has its adjoint $\widetilde{\mathcal{L}}^*$ $=$
$\rho^{-1}(x)\nabla\big( A(x)\rho(x)\nabla\big)$ $-j(x)\nabla$, where
the $j(x)$ is defined in Eq. (\ref{aode}) below.  The present
paper chooses to work with the forward operator $\mathcal{L}$ rather
than the backward operator $\widetilde{\mathcal{L}}$.

\section{Symmetric reversible diffusion}

The multi-dimensional elliptic differential operator
\begin{equation}
  \mathcal{L}_s\left(u\right)
          = \nabla\Big(A(x)\nabla u(x) 
             - \big(A(x)\nabla\ln\rho(x)\big) 
                          u(x)\Big)
\label{sfpe}
\end{equation}
is a Fokker-Planck type with diffusion tensor $A(x)$ and drift
$A(x)\nabla\ln\rho(x)$.  One can immediately check that $\rho(x)$ 
is its stationary density.  The diffusion process associated 
with $\mathcal{L}_s$ has been extensively studied in the past
and is well understood.

The corresponding partial differential equation 
(PDE) 
\begin{equation}
    \frac{\partial u(x,t)}{\partial t}
           = \mathcal{L}_s\big(u(x,t)\big)
\label{symmetric_pde}
\end{equation}
is formally a gradient system
\begin{equation}
           \frac{\partial u}{\partial t}
         = \frac{\delta}{\delta u}
                \frac{\langle \mathcal{L}_s(u),u\rangle}{2}.
\label{grad_flow}
\end{equation}
See \cite{qqt_jsp} for many important properties 
associated with reversible diffusion processes.   

{\bf\em Potential function, entropy production and stochastic
generalized free energy.} The ``potential function'' in 
Eq. (\ref{grad_flow}) can be expressed as 
\begin{equation}
   -\langle\mathcal{L}_s(u),u\rangle =
   \int_{\mathbb{R}^n} \left[\nabla\left(\frac{u(x,t)}{\rho(x)}
          \right)\right]^T A(x) \left[ \nabla\ln
         \left(\frac{u(x,t)}{\rho(x)}
          \right)\right] u(x,t)dx.
\end{equation}
The gradient system in (\ref{grad_flow}) is confined on an
affine subspace with $\int_{\mathbb{R}^n} u(x)dx$ $=$
$\langle \rho(x),u(x)\rangle$ $= 1$.
The PDE (\ref{symmetric_pde}) has also been shown as a gradient 
flow generated by a potential function $F[u]$ on an 
appropriate Riemann manifold with Wasserstein metric
\cite{otto,chow}.  The $F[u]$ is known as  
the ``stochastic generalized free energy'' for the Markov system
\cite{mackey,ao_ctp,ge_qian_10,3faces}: 
\begin{equation}
      F\big[u(x,t)\big] = \int_{\mathbb{R}^n} u(x,t)
         \ln\left(\frac{u(x,t)}{\rho(x)}\right)dx  
           = \big\langle \rho u,\ln\left(\rho^{-1}u\right)
                \big\rangle. 
\end{equation}
Then for $u(x,t)$ following the self-adjoint
PDE (\ref{symmetric_pde}) \cite{spohn,ge_qian_10,3faces}, 
\begin{eqnarray}
  \frac{dF(t)}{dt} &=& \Big\langle \rho 
                \frac{\partial u}{\partial t},\ln\left(\rho^{-1}u\right)
                \Big\rangle +
         \Big\langle \rho u, u^{-1}\frac{\partial u}{\partial t}
                \Big\rangle
\nonumber\\
   &=& \Big\langle \mathcal{L}_s(u),
                \rho\ln\left(\rho^{-1}u\right)
                \Big\rangle \ = \ -e_p.
\label{def_epr}
\end{eqnarray}
The $e_p\ge 0$ is known as the entropy production rate for the 
reversible diffusion process \cite{qqt_jsp,jqq}.  A stationary 
reversible diffusion, therefore, has $e_p=0$.

{\bf\em Stochastic differential equation for trajectories.}
One can also write a stochastic differential equation 
for the diffusion process described by Eq. (\ref{symmetric_pde}).  If 
we follow It\={o}'s notion of stochastic integration, we have
\begin{equation}
   dx_i(t) = \rho^{-1}(\vec{x})\sum_{j}\left(
       \frac{\partial}{\partial x_j} A_{ij}(\vec{x})\rho(\vec{x})
                \right)dt + \sum_{j} \Gamma_{ij}(\vec{x})dB_j(t),
\end{equation}
in which matrix 
$A(\vec{x})=\frac{1}{2}\Gamma(\vec{x})\Gamma^T(\vec{x})$, 
and $B_i(t)$ are standard Brownian motions.
On the other hand, if one follows Stratonovich's 
integration, one has
\begin{equation}
   dx_i(t) = \sum_j \left(A_{ij}(\vec{x})
         \frac{\partial\ln\rho(\vec{x})}{\partial x_j}
         +\frac{1}{2}\sum_k \Gamma_{ik}(\vec{x})
         \frac{\partial}{\partial x_j}\Gamma_{jk}(\vec{x})
         \right)dt + \sum_{j} \Gamma_{ij}(\vec{x}) dB_j(t),
\end{equation}
and if one takes the ``divergence form'' for the 
integration, as strongly advocated by P. Ao \cite{ao_jpa_04,aoping_qian}, then
\begin{equation}
  dx_i(t) = \sum_{j}\left(A_{ij}(\vec{x})
       \frac{\partial}{\partial x_j}\ln\rho(\vec{x})
                \right)dt + \sum_{j} \Gamma_{ij}(\vec{x})dB_j(t).
\end{equation}

\section{Canonical conservative dynamics}

On the other hand, the anti-symmetric partial differential 
operator is
\begin{eqnarray}
 \mathcal{L}_a\left(u\right) &=& \nabla\Big(\big(
               A(x)\nabla\ln\rho(x)-b(x)\big) u(x)\Big)
\\
     &=& \big(A(x)\nabla\rho(x)-b(x)\rho(x)\big)\cdot
             \nabla\big(\rho^{-1}(x)u(x)\big).
\label{afpe}
\end{eqnarray}
The corresponding PDE 
\begin{equation}
      \frac{\partial u(x,t)}{\partial t} =
                   \mathcal{L}_a\big(u(x,t)\big)
\label{a_pde}
\end{equation}
is not diffusive but rather it is a first-order hyperbolic PDE.  It is 
easy to verify that $\rho(x)$ is again a stationary density for 
Eq. (\ref{a_pde}).  Eq. (\ref{afpe}) actually is the Liouville 
equation in the phase space for the nonlinear ordinary 
differential equation (ODE)
\begin{equation}
    \frac{dx}{dt} = b(x)-A(x)\nabla\ln\rho(x)\equiv 
             j(x).
\label{aode}
\end{equation}
The $j(x)$ in Eq. (\ref{aode}) satisfies 
\begin{equation}
  \nabla\cdot\big(\rho(x) j(x) \big) = 0.
\label{div_free}
\end{equation}
That is,
\begin{equation}
      \nabla\cdot j(x)+j(x)\cdot\nabla\ln\rho(x) = 0.
\label{eq_xyz}
\end{equation}
Theresore, Eq. (\ref{aode}) is a canonical conservative system 
with respect to stationary density $\rho(x)$.
Since $\rho(x)$ is an invariant density to the dynamics in 
Eq. (\ref{a_pde}), one can again consider the generalized free 
energy functional \cite{mackey,ao_ctp,ge_qian_10,3faces}
\begin{equation}
       F\big[u(x,t)\big] = \int_{\mathbb{R}^n}
       u(x,t)\ln\left(\frac{u(x,t)}{\rho(x)}\right)
                  dx.
\end{equation}
For the hyperbolic system, this is a generalization 
of Boltzmann's $H$-function in which the stationary 
distribution $\rho(x)=$ constant due to 
``equal probability a priori''.  Then one has
\cite{voigt,mackey}
\begin{eqnarray}
   \frac{dF(t)}{dt} &=& \int_{\mathbb{R}^n}
                \frac{\partial u(x,t)}{\partial t}
           \ln\left(\frac{u(x,t)}{\rho(x)}\right) dx
\nonumber\\
        &=& \int_{\mathbb{R}^n}
                j(x)u(x,t)\left[
            \frac{\nabla u(x,t)}{u(x,t)}
              -\frac{\nabla\rho(x)}{\rho(x)}
          \right] dx
\nonumber\\
          &=& -\int_{\mathbb{R}^n}
                \big(\nabla\cdot j(x)+ 
               j(x)\cdot \nabla\ln\rho(x)
               \big)u(x,t)\ dx 
\nonumber\\
             &=& 0.        
\label{dfdtiszero}   
\end{eqnarray}
In the derivation we have used Eqs. (\ref{a_pde}) and
(\ref{eq_xyz}) as well as assumed that $u(x,t)\rightarrow 0$ 
sufficiently fast when $\|x\|\rightarrow\infty$.

The result in Eq. (\ref{dfdtiszero}) is reminiscent of a 
Boltzmann's result before he introduced his {\em Stosszahlansatz}.
His $H$-function, $-\int f(x,t)\ln f(x,t)\ dx$, actually is 
invariant with respect to time if $f(x,t)$ follows 
strictly the phase space Liouville equation for a Hamiltonian 
dynamics \cite{dorfman}.  One also notes that the $L_2$ norm of $u(x,t)$ is
conserved in a canonical conservative dynamics:
$\frac{d}{dt}\|u(x,t)\|^2 =$ $\big\langle\mathcal{L}_a(u),u\big\rangle =$
$-\big\langle u,\mathcal{L}_a(u)\big\rangle = 0$.
Actually, any functional $\int_{\mathbb{R}^n} uG(u/\rho)dx$ 
is a conserved quantity for Eq. (\ref{a_pde}).

{\bf\em Fixed points in canonical conservative system.} 
Eq. (\ref{eq_xyz}) indicates that $\nabla\cdot j(x)=0$ 
at the fixed point of vector field $j(x)=0$. 
Without loss of generality, let $x^*=0$ be a fixed 
point: $j(x^*)=0$.  Then
in the neighbourhood of $x^*=0$ one has
\begin{equation}
     j(x) = Bx + \frac{1}{2}x^TGx + \cdots, \ \
     \rho(x) = \rho(0) + \vec{q}\cdot x + \cdots,
\end{equation}
in which the Jacobian matrix $B$ has elements $B_{\ell k} =
\partial j_{\ell}(0)/\partial x_k$, tensor elements 
$G_{\ell kh} = \partial^2 j_{\ell}(0)/\partial x_k\partial x_h$,
and gradient
$\vec{q}$ has elements $q_k=\partial\rho(0)/\partial x_k$.
Then the Eq. (\ref{eq_xyz}) becomes
\begin{equation}
     \rho(0)\textrm{Tr}[B] + \vec{q}Bx 
   + \textrm{Tr}[B]\vec{q}\cdot x + \frac{\rho(0)}{2}
    \sum_{\ell k}\left\{G_{\ell\ell k}x_k+
             G_{\ell k\ell}x_k  \right\}
         + O\left(\|x\|^2\right) = 0,
\end{equation}
$\forall x$. Since $\rho(0)\neq 0$ this yields Tr$[B]=0$.
Furthermore, 
\begin{equation}
        \sum_{\ell} \frac{\partial\ln\rho(0)}{\partial x_{\ell}} 
       \frac{\partial j_{\ell}(0)}{\partial x_k} + 
       \frac{\partial^2 j_{\ell}(0)}
     {\partial x_{\ell}\partial x_k}  = 0.
\end{equation}
Hence, the fixed point of a canonical conservative system can 
not be a node or focus.  It has to be either a saddle or a 
center, just as in a Hamiltonian system \cite{perko}.

{\bf\em Stationary points of \boldmath{$\rho(x)$} in canonical conservative system.}  Eq. (\ref{eq_xyz}) also indicates that
$\nabla\cdot j(x)=0$ at the stationary position of $\rho$ 
where $\nabla\rho(x^s)=0$.  A similar local analysis can 
be carried out near $x^s$, and one has
\begin{equation}
    \sum_{\ell} j_{\ell}(x^s)\frac{\partial^2\rho(x^s)}
        {\partial x_{\ell}\partial x_k} + 
    \frac{\partial^2 j_{\ell}(x^s)}
          {\partial x_{\ell}\partial x_k} = 0.
\end{equation}

{\bf\em Planar canonical conservative system.} A
planar canonical conservative system has the general form 
$\dot{x}=\rho^{-1}(x,y)\partial H(x,y)/\partial y$, 
$\dot{y}=-\rho^{-1}(x,y)\partial H(x,y)/\partial x$.
The phase portrait for this system is identical
to the Hamiltonian system with $\rho=1$.  Indeed,
$H(x,y)$ is a conserved quantity in the dynamics:
$dH(x(t),y(t))/dt = 0$.

{\bf\em Mapping to microcanonical conservative system via time change.} 
More generally, let $\rho(x)j(x)=f(x)$.  Let $\hat{x}(t)$
be a solution to the microcanonical conservative system 
$\dot{x}=f(x)$.  Then the solution to the canonical 
conservative system with same initial value is 
$x(t)=\hat{x}\left(\hat{t}(t)\right)$
in which
\[      \hat{t}(t) = t_0+\int_{t_0}^t
           \rho^{-1}\left(\hat{x}(s)\right)ds.
\]

\section{General diffusion processes without detailed
balance}

We now bring the results from the above two sections 
to bear on the general diffusion processes. 

{\bf\em Decomposing free energy dissipation \boldmath{$\frac{dF}{dt}$}.} 
In terms of $j(x)$, the ``thermodynamic force'' \cite{onsager} in a
general diffusion generated by $\mathcal{L}$ \cite{qqt_jsp} 
\begin{equation}
    b(x) - A(x)\nabla\ln u(x) = j(x) -
         \Big\{ A(x)\nabla\ln\left(\frac{u(x)}{\rho(x)}
            \right) \Big\},
\label{eq_26}
\end{equation}
in which the $j(x)$ characterizes the deviation of
$b(x)$ from a gradient force; it does not involve $u(x)$.
And since the $\rho(x)$ is uniquely determined by 
the diffusion matrix and drift, $j(x)$ is a stationary 
term strictly determined by the $A(x)$ and $b(x)$.  
The term in $\{\cdots\}$ characterizes non-stationarity 
of the $u(x)$.  Then we have, 
\begin{equation}
  \Big\langle \mathcal{L}(u),
              \rho\ln\left(\rho^{-1}u\right) \Big\rangle
      = \Big\langle \mathcal{L}_s(u),
              \rho\ln\left(\rho^{-1}u\right) \Big\rangle
      + \Big\langle \mathcal{L}_a(u),
              \rho\ln\left(\rho^{-1}u\right) \Big\rangle.
\end{equation}
The last term is zero for the canonical 
conservative system, e.g., Eq. (\ref{dfdtiszero}).
Therefore, the $\frac{dF}{dt}$ for a general diffusion 
process with $\mathcal{L}$ is entirely due to its symmetric 
part of the diffusion, $\mathcal{L}_s$.  The canonical
conservative dynamics generated by $\mathcal{L}_a$
has no contribution toward the {\em generalized free energy dissipation}
of $\mathcal{L}$.

{\bf\em Non-negative source for generalized free energy 
\boldmath{$F[u]$}.}  Another important quantity from 
physics, the {\em house-keeping heat} first 
proposed by Oono and Paniconi \cite{oono,ge_qian_10},
also called adiabatic entropy production \cite{3faces,3faces_pre,sasa01}, 
is
\begin{equation}
       E_{in}=\int_{\mathbb{R}^n}
          \big(b-A\nabla\ln\rho\big)A^{-1}(x)
          \big(bu-A\nabla u\big) dx =
    \Big\langle \rho j,
      A^{-1}\big(bu-A\nabla u\big) 
      \Big\rangle   
\end{equation}
for a general diffusion process.  It is a type 
of projection of the thermodynamic driving force 
$\big(b(x)-A(x)\nabla\ln u(x)\big)$ onto the
$j(x)$.  It is zero for a reversible diffusion 
process with $j(x)=0$. Noting the Eq. (\ref{eq_26}), 
we have
\begin{equation}
       E_{in} = \Big\langle \rho j,
      A^{-1}u j \Big\rangle -
      \Big\langle \rho j,
      \rho\nabla\big(\rho^{-1}u\big) 
      \Big\rangle = \Big\langle \rho j,
      uA^{-1} j \Big\rangle \ge 0.
\end{equation}
Then the free energy dissipation for a general diffusion
\begin{equation}
           \frac{dF(t)}{dt} = E_{in}(t) - e_p(t),
          \textrm{ or } 
          e_p(t) = E_{in}(t) - \frac{dF(t)}{dt}.
\label{main}
\end{equation}
All three terms $e_p$, $E_{in}$, $-\frac{dF}{dt}$ 
are non-negative.  A symmetric
diffusion has $E_{in}(t)=0$ $\forall t$ (Eq. \ref{def_epr});
a canonical conservative dynamics has $\frac{dF(t)}{dt}=0$
$\forall t$ (Eq. \ref{dfdtiszero}).  A general 
diffusion process without detailed balance has $e_p(t)$
which is consist of non-negative $-\frac{dF}{dt}$ 
and $E_{in}$.  Logically, we believe the mathematical Eq. (\ref{main})
should be {\em interpreted} as a balance equation for the generalized
free energy $F[u]$ with a source $E_{in}$ and a sink $e_p$.
The fact that both $E_{in}$ and $e_p$ are non-negative
indicates that they represent the authentic source and sink terms
for the $F[u]$.

\section{Ornstein-Uhlenbeck process: the linear case}

	We now consider the relationship between the above 
result and P. Ao's decomposition \cite{ao_jpa_04}.
We shall explicitly work out the details for the 
Ornstein-Uhlenbeck (OU) Gaussian process 
\cite{qian_prsa,kat}.  We show that in the 
linear case, the two theories are equivalent.  The
present approach is also equivalent to that of
R. Graham's  \cite{graham_tel} under the assumption
of large deviation principle.

	We consider linear vector field $b(x)=Bx$ and
constant diffusion matrix $A$.  The $n$-dimensional
Fokker-Planck equation is 
\begin{equation}
    \frac{\partial u(x,t)}{\partial t}
         = \nabla \big(A\nabla u(x,t)-Bx u(x,t)\big).
\end{equation}
It is easy to verify that the stationary solution
has a Gaussian form \cite{qian_prsa}
\begin{equation}
     \rho(x) = \frac{1}{(2\pi)^{n/2}\det\Xi}
                \exp\left\{-\frac{1}{2} x^T\Xi^{-1}x  
            \right\},
\label{gaussian}
\end{equation}
in which the covariant matrix $\Xi$ satisfies the
Lyapunov matrix equation
\begin{equation}
          B\Xi+\Xi B^T + 2A = 0.
\label{eq4xi}
\end{equation}    
Accordingly, we have
\begin{equation}
        Bx = A\nabla\ln\rho(x)+j(x)= -A\Xi^{-1}x+j(x) , \ \
        j(x) = Jx,
\end{equation} 
where matrix $J=B+A\Xi^{-1}$.  We now show that matrix $J$
can be written as $-R\Xi^{-1}$ in which $R$ is an
anti-symmetric matrix.  This is because of Eq. (\ref{eq4xi}), 
\[
    -R^T  = (B\Xi+A)^T = \Xi B^T+A = -(B\Xi+A)=R.
\]
Therefore, $B=-(A+R)\Xi^{-1}$.  So the linear stochastic
differential equation $dx(t) = Bxdt + \Gamma dB(t)$, with
$\frac{1}{2}\Gamma\Gamma^T=A$, can be re-written as 
\begin{eqnarray}
   M dx(t) &=& -\nabla\left(\frac{1}{2}x^T\Xi^{-1}x\right) dt 
                      + \Pi dB(t)
\nonumber\\
           &=& \nabla\left(\ln\rho(x)\right) dt + \Pi dB(t),
\label{aoping1}
\end{eqnarray}
where $\Pi=(A+R)^{-1}\Gamma$ and 
$M=(A+R)^{-1}$.  They are related via
\begin{equation} 
      M+M^T = \Pi\Pi^T,
\label{aoping2} 
\end{equation}
because 
\[
    \Pi^{-1}\left[(A+R)^{-1}+(A+R)^{-T}\right]\Pi^{-T}
  = \Gamma^{-1}\left[(A-R)+(A+R)\right]\Gamma^{-T}
\]
\[
  = 2\Gamma^{-1}A\Gamma^{-T} 
  = I.
\]
Eq. (\ref{aoping1}), together with (\ref{aoping2}), 
is Ao's form of stochastic differential 
equation \cite{ao_jpa_04}.
For the linear system, one also has an additional property:
the gradient field $-\Xi^{-1} x$ and canonical conservative 
dynamics $Jx$ are actually orthogonal:
\begin{equation}
      \left(\Xi^{-1}x\right)^T\cdot (J x) 
         = -x^T\Xi^{-1}R\Xi^{-1}x
         =-\left(\Xi^{-1}x\right)^T R \left(\Xi^{-1}x\right)
         = 0.
\end{equation}
This orthogonality was noted in both \cite{graham_tel}
and \cite{kat}. See \cite{qian_prsa} and \cite{kat} for 
more discussions on irreversible OU processes.

\section{Discussion}

The above result provides some insights into the
structural stability of non-gradient vector field $b(x)$.
Zeeman \cite{zeeman} has advocated an approach to structural 
stability based on $\epsilon$-noise perturbed dynamical
systems.  With respect to non-gradient field with a focus,
he clearly noted one key difficulty in the Morse-Smale 
theory which requires a mapping from a focus to a node.  
While such a homeomorphism exists, ``It is impossible,
however, to make [such a map] smooth. ... 
Therefore in the attempt to capture 
density in two dimensions we have to abandon smoothness in 
the very definition of structural stability, and this, alas, 
is the beginning of the rot.'' \cite{zeeman}

For simplicity, let $A(x)=\epsilon$, then 
$b(x)=j_{\epsilon}(x)+\epsilon\nabla\ln\rho_{\epsilon}(x)$ 
in which $\rho_{\epsilon}(x)$ is the stationary density for
the randomly perturbed dynamical system $dx(t)=b(x)dt+\epsilon dB(t)$, and $j_{\epsilon}(x)$ is a canonical conservative 
system.  In the limit of $\epsilon\rightarrow 0$, whether
a limit exists for $-\epsilon\ln\rho_{\epsilon}(x)$
is precisely the theory of large deviations \cite{fw_book,ge_qian_HJE}.  If a differentiable limit 
$U(x)$ exists, then one obtains a decomposition 
$b(x)=j_0(x)-\nabla U(x)$.  Moreover, since 
$\nabla\cdot\left(\rho_{\epsilon}(x)j_{\epsilon}(x)\right)=0$
$\forall\epsilon$, one also has 
$-\epsilon j_{\epsilon}(x)\cdot\nabla\ln\rho_{\epsilon}(x)$ $=$ $\epsilon\nabla\cdot j_{\epsilon}(x)$.  Thus
$j_0(x)\cdot\nabla U(x)=0$ if the convergence is uniform, i.e.,
the decomposition is orthogonal. This is indeed R. Graham's
theory \cite{graham_tel,ge_qian_HJE}. 
On the other hand, if $\rho_{\epsilon}(x)$ has 
a nonzero limiting density $\rho_0(x)$ on an attractor, 
then $U(x)=0$ on the entire attractor.  This has been 
explicitly shown for systems with limit cycles and 
invariant tori where the asymptotic stationary density
$\rho_{\epsilon}(x) \sim \rho_0(x)e^{-U(x)/\epsilon}$
\cite{ge_qian_chaos}. 
However, if the limit $U(x)$ is non-differentiable or worse, 
some weaker forms might still exist \cite{lsyoung}. This
is the technical challenges encountered by R. Graham and 
his coworkers \cite{graham_tel}; further investigations 
are required.  The present work could provide a different
approach to the challenge.

	Canonical conservative system with 
$\frac{dF(t)}{dt}=E_{in}(t)-e_p(t)=0$ also leads to an 
interesting contradistinction between two views on time irreversibility.  It is now known that entropy production
$e_p$ can be defined as the relative entropy between the
probabilities of a trajectory $\omega_t$ and its time 
reversal $r(\omega_t)$, under the probability measure generated by
$\mathcal{L}$ \cite{qq85,seifert_05,ge_jiang}. 
For a stationary diffusion, the probability for the $r(\omega_t)$ with the 
measure generated by $\mathcal{L}=\mathcal{L}_s+\mathcal{L}_a$ 
is in fact the same as the $\omega_t$ under the measure 
generated by $\mathcal{L}^-=\mathcal{L}_s-\mathcal{L}_a$ 
\cite{ge_jiang}. Now, if one introduces a different
entropy production $e_p^{\#}$ as \cite{3faces,3faces_pre}
\begin{equation}
    e^{\#}_p = E^{\mathbb{P}}\left[
          \ln\left(\frac{d\mathbb{P}^{\ \ }}
          {d\mathbb{P}^{-\#}}(\omega_t)\right)\right],
\end{equation}
in which the $\mathbb{P}^{-\#}$ is defined as 
the probability of time-reversed trajectory $r(\omega_t)$
under the measure generated by $\mathcal{L}^-$. Then
$\mathbb{P}^{-\#}=\mathbb{P}$ and $e_p^{\#}=0$ for any
stationary diffusion, as well as the canonical
conservative dynamics. With such a choice for {\em the
definition of time reversal}, there will be no
nonequilibrium steady state; only equilibrium in
which $e_p^{\#}=E_{in}^{\#}=0$.  Indeed, a Hamiltonian system 
is considered to be time reversible in classical dynamics
precisely due to the second type of time reversal \cite{kim_qian_prl}.  
Under $e_p$, cycle kinetics is interpreted as a driven phenomenon;
under $e_p^{\#}$, cyclic dynamics is explained as a consequence
of ``inertia''.  Indeed, a map between nonequilibrium
steady state and Hamiltonian system with rotation
has also been proposed in \cite{xing}.

Finally, we note that the decomposition of a general
diffusion process into an $\mathcal{L}_a$ and an
$\mathcal{L}_s$ parts unifies nicely the earlier 
mathematical theories of dynamics formulated respectively 
by Newton and Fourier \cite{newton,fourier}.  We also
note, with cosmological dynamics in mind, that in
a long-time limit, the $\mathcal{L}_s$ part of a dynamics
vanishes, only the $\mathcal{L}_a$ part remains
with permanence.

\vskip 0.5cm\noindent

I thank Ping Ao, Zhen-Qing Chen, Hao Ge, Da-Quan Jiang, Yao Li, 
Jin Wang, and Jianhua Xing for many helpful discussions.  I 
particularly acknowledge Dr. P. Ao for his inspiration, 
encouragement, and comradeship.  The present work
grew out of a desire to reconcile the decomposition in 
his Darwinian dynamics \cite{ao_ctp} and the nonequilibrium
steady-state circulation decomposition theorems in \cite{qq82,jqq}.

\end{document}